\documentclass{vgtc}                          

\usepackage[subtle]{savetrees}
\ifpdf
  \pdfoutput=1\relax                   
  \usepackage{graphicx}                
  \DeclareGraphicsExtensions{.pdf,.png,.jpg,.jpeg} 
\else
  \usepackage{graphicx}                
  \DeclareGraphicsExtensions{.eps}     
\fi%

\graphicspath{{figures/}{pictures/}{images/}{./}} 

\usepackage{microtype}                 
\PassOptionsToPackage{warn}{textcomp}  
\usepackage{textcomp}                  
\usepackage{mathptmx}                  
\usepackage{times}                     
\usepackage{cite}                      
\usepackage{tabu}                      
\usepackage{booktabs}                  

\usepackage{xcolor}
\usepackage{amsmath}
\usepackage{float}
\usepackage{dblfloatfix}
\usepackage[colorlinks=true,linkcolor=blue,urlcolor=blue,citecolor=blue]{hyperref}

\onlineid{0}

\vgtccategory{Research}

\vgtcinsertpkg

\title{\textit{n} Walks in the Fictional Woods}

\usepackage{tikz}

\usetikzlibrary{svg.path}

\definecolor{orcidlogocol}{HTML}{A6CE39}

\tikzset{
orcidlogo/.pic={
    \fill[orcidlogocol] svg{M256,128c0,70.7-57.3,128-128,128C57.3,256,0,198.7,0,128C0,57.3,57.3,0,128,0C198.7,0,256,57.3,256,128z};
    \fill[white] svg{M86.3,186.2H70.9V79.1h15.4v48.4V186.2z}
    svg{M108.9,79.1h41.6c39.6,0,57,28.3,57,53.6c0,27.5-21.5,53.6-56.8,53.6h-41.8V79.1z M124.3,172.4h24.5c34.9,0,42.9-26.5,42.9-39.7c0-21.5-13.7-39.7-43.7-39.7h-23.7V172.4z}
    svg{M88.7,56.8c0,5.5-4.5,10.1-10.1,10.1c-5.6,0-10.1-4.6-10.1-10.1c0-5.6,4.5-10.1,10.1-10.1C84.2,46.7,88.7,51.3,88.7,56.8z};
  }
}

\newcommand{\orcid}[1]{%
\hypersetup{urlcolor=black}%
\href{https://orcid.org/#1}{\mbox{{
      \begin{tikzpicture}[yscale=-1,transform shape, scale=0.18, every node/.style={scale=0.18}]%
        \pic{orcidlogo};%
      \end{tikzpicture}%
    }}
    }%
}

\author{
\textcolor{gray}{main:}\\
Victor Schetinger \orcid{0000-0002-8116-794X} \thanks{e-mail: lbschetinger@fhstp.ac.at}\\ %
        \scriptsize{FH St. Pölten, Austria} %
\and 
    \textcolor{gray}{equal contribution:}
    $
    \begin{cases}
      & \text{Sara Di Bartolomeo \orcid{0000-0001-9517-3526} --- \scriptsize{Universität Konstanz, Germany}}\\
      & \text{Edirlei Soares de Lima \orcid{0000-0002-2617-3394} --- \scriptsize{Universidade Europeia, Portugal} }\\
      & \text{Christofer Meinecke \orcid{0000-0002-5637-9975} --- \scriptsize{Leipzig University, Germany}}\\
      & \text{Rudolf Rosa \orcid{0000-0003-4908-6127} --- \scriptsize{Charles University, Czechia}}
    \end{cases}  
    $
}

\abstract{This paper presents a novel exploration of the interaction between generative AI models, visualization, and narrative generation processes, using OpenAI's GPT as a case study. We look at the question \textbf{``Where Does Generativeness Comes From''}, which has a simple answer at the intersection of many domains. Drawing on Umberto Eco's ``Six Walks in the Fictional Woods'', we engender a speculative, transdisciplinary scientific narrative using ChatGPT in different roles: as an information repository, a ghost writer, a scientific coach, among others. The paper is written as a piling of plateaus where the titling of each (sub-)section, the ``teaser'' images, the headers, and a biblock of text are strata forming a narrative about narratives. To enrich our exposition, we present a visualization prototype to analyze storyboarded narratives, and extensive conversations with ChatGPT. Each link to a ChatGPT conversation is an experiment on writing where we try to use different plugins and techniques to investigate the topics that, ultimately form the content of this portable document file. Our visualization uses a dataset of stories with scene descriptions, textual descriptions of scenes (both generated by ChatGPT), and images (generated by Stable Diffusion using scene descriptions as prompts). We employ a simple graph-node diagram to try to make a ``forest of narratives'' visible, an example of a vis4gen application that can be used to analyze the output of Large Languange + Image Models.

} 

\teaser{
  \centering
  \includegraphics[width=\linewidth]{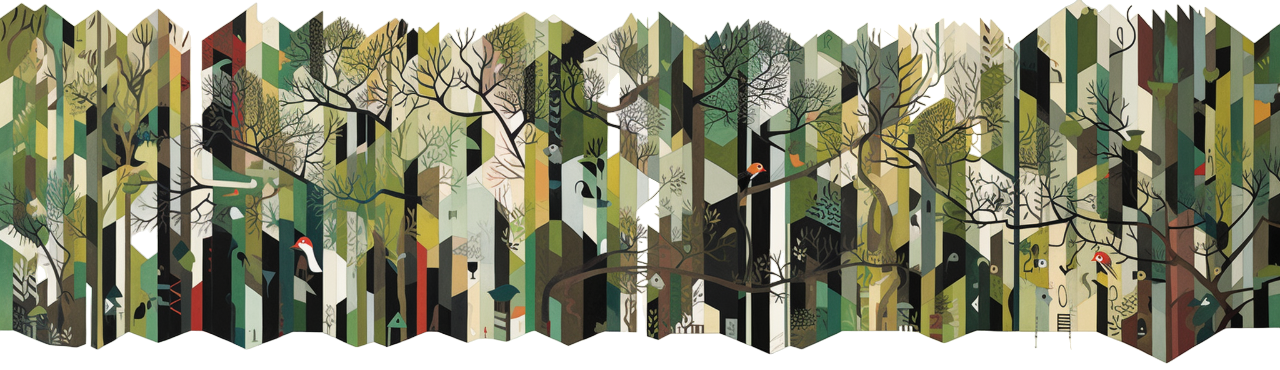}
  \label{fig:teaser}
}

\begin{document}



\maketitle

\section*{Prologue}



In ``Six Walks in the Fictional Woods"~\cite{8ab239b9-bfbc-35a3-8699-8e27087515dd}, Umberto Eco discusses how the interplay between author, text, and reader can generate infinite combinatorial spaces. Rather than merely parroting tired truisms such as \textit{``the reader brings himself to the text"} or \textit{``personal experience impacts interpretation"}, Eco delves deeply into the vast expanse of cognitive constructs formed within the minds of readers while interacting with a text (i.e., what is being \textit{rendered} in the mind as one reads). Eco's foresight is evident in his acknowledgement of the potential role of AI in influencing this triadic process, a topic he touches upon multiple times in his books. The plot of ``Foucalt's Pendulum"~\cite{eco1990foucaults}, for example, involves a computer (Abulafia) that is used to generate fictions, and these fictions start affecting reality in unexpected ways.

Eco's metaphorical woods, \textit{``...tangled and twisted like the forests of the Druids, and not orderly like a French garden.''} [sic], encapsulate the multidimensional, combinatorial spaces formed by the reading process of humans and machines. His six walks provide insights into varying facets of textual interaction, culminating in a discussion of ``Fictional Protocols". Though the term `generative' is not explicitly employed, this chapter essentially discusses where `generativeness' in text comes from.

\begin{figure*}[ht!]
    \vspace*{-2.55cm}
    \hspace{-3.9cm}
    \includegraphics[width=1.4\textwidth]{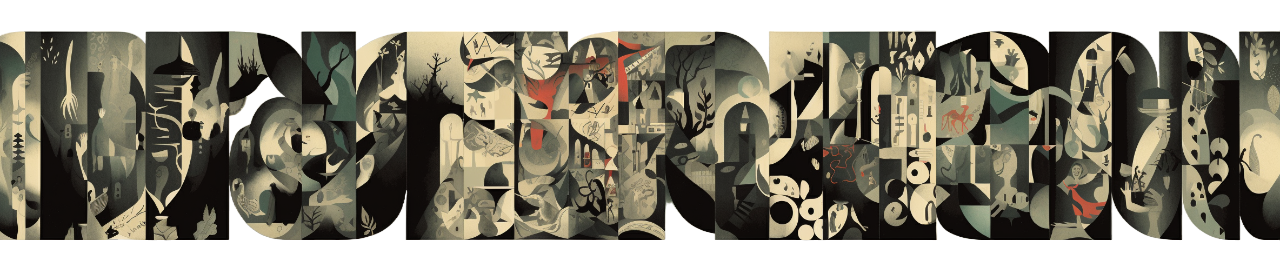}
\end{figure*}
\subsection*{Where Does Generativeness Comes From?}

\includegraphics[width=\linewidth]{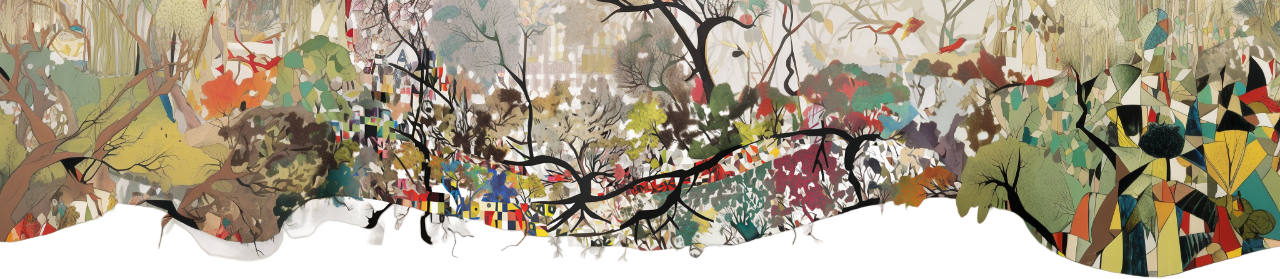}

In short, `generativeness' comes from the existence of latent spaces organized around \textit{meaningful} principles, which reality seems somehow to be plentiful of. A literary text, according to Eco, organizes a latent space around it (the fictional \textit{woods}) based on its semiotic ``hyperstructure'', that is, its possible relationships with reality and potential readers' sign systems. While the philosopher Heraclitus says that \textcolor{blue}{\href{https://chat.openai.com/share/77edeaa8-20d7-4cbc-af93-ca69c76e770c}{you cannot step into the same river twice}}, Eco also says the same of the fictional woods. The second time a reader reads a text, the mental constructs invoked (and rendered) in the mind will be affected by the first reading, and so on. Can we then say that every text is a sort of analogical generative model that, when coupled with a reader can produce infinite mental landscapes?

Let us start by recognizing that humans are not equipped to deal with infinity. We can symbolically manipulate it, engineer it into useful things in the physical world, but we almost never really have face-to-face encounters with it. Those who learned calculus and struggled with the concept of limits realize there is always a Zeno-like leap of faith from infinitesimally surfing a curve to landing on a point. We are constantly drowning in a sea of infinities, and every thing we can actually experience or think is just a {\tiny tiny,} {\small \textit{small},} subset of {\large \textbf{all}} possibilities. Therefore, `generativeness' is a ruse, generative models are less like factories producing new things and more like telescopes (or microscopes, or MRIs) pointing at unexplored places in vastly infinite latent spaces.


\subsection*{O(n) the Scale of Human Potential}

\includegraphics[width=\linewidth]{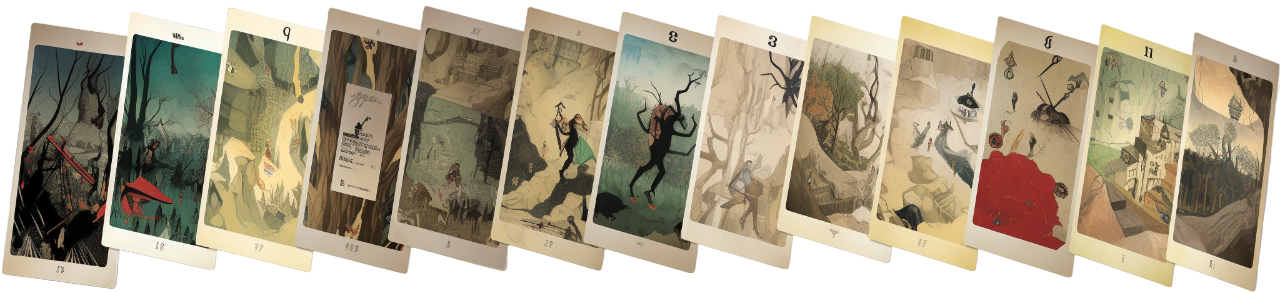}

A single human life, in seconds, is around \(2.29 \times 10^9\) seconds. The estimated total number of humans (Homo Sapiens) who have ever lived is approximately \(108.5 \times 10^9\). Therefore, the total number of seconds that all humans have lived, considering the average lifespan and the estimated total number of humans who have ever lived, is approximately \(2.48 \times 10^{20}\) seconds. While this might look like a large number, any programmer will tell you its not, that these are rookie numbers. A badly coded sorting algorithm could easily beat this in terms of running time. If all humans coordinated to count all natural numbers, one (or ten, or a hundred) per second since the beginning of time, we would not be too far from the start. If we are talking about real numbers, we would not have arrived at 0.1. 

In comparison, the number of ways to shuffle a standard deck of 52 cards is approximately \(8.07 \times 10^{67}\). This is an astronomically larger number, much bigger than the estimated amount of atoms in the universe, demonstrating that even the vast span of human existence is minuscule in comparison to the combinatorial possibilities of something as simple as shuffling a deck of cards. If we devoted ourselves to the task of shuffling all possible combinations instead of counting numbers, we would be doing even worse. Yet, humanity keeps tapping itself on the back for all its amazing achievements, alone in the universe with the burden of intelligence. How come? How could we have done this with such limited time, specially considering we have to split our time between the physical and mental worlds?

\section*{What the World Affords Us}
\includegraphics[width=\linewidth]{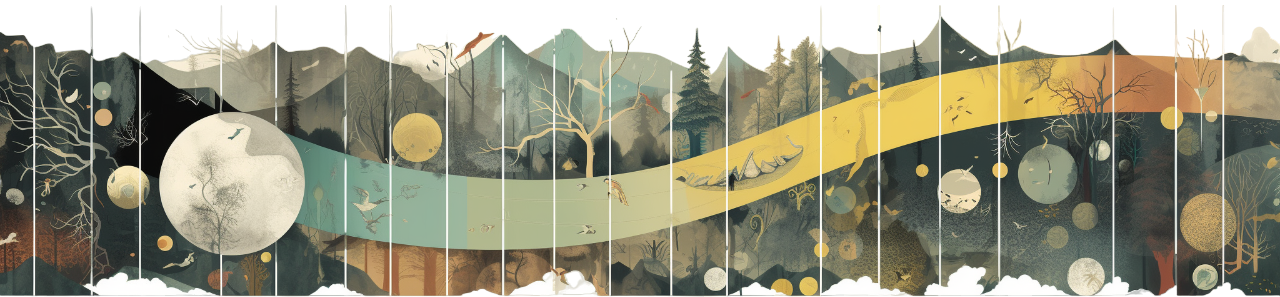}

In his book ``Materialist Phenomenology: A Philosophy of Perception"~\cite{delanda_materialist_nodate}, DeLanda provides clues to answer this question through a comprehensive exploration of the synthesis of the visual field, which he argues is a key component of our understanding and interaction with the world. He proposes a non-reductive materialist approach, which asserts that there are mental properties that are different from physical properties, that the existence of mental properties depends on the existence of physical properties, and that mental properties can confer causal powers on mental events. He poses that the world affords us structures, and that humans can heuristically attach themselves to the combinatorial spaces around these structures to effectively navigate reality despite our limited asymptotic time. The consistent physical behavior of things around us, having a ground under our feet, a sky over our head with stars, a moon, and a sun, act as constants that allow us to reduce our cognitive search spaces. In trying to minimize \textit{surprise}~\footnote{\textcolor{blue}{\href{https://www.youtube.com/watch?v=jZ1fsXQz7M4&t=585s}{https://www.youtube.com/watch?v=jZ1fsXQz7M4\&t=585s}}}in this environment, we develop our own generative models of it.


DeLanda's arguments are rooted in the belief that there are entities which are independent of the existence of our minds, such as the geological, climatological, and ecological processes that shaped the planet on which we evolved; and there are entities that are independent of the content of our minds, that is, entities that have a definite nature which does not change when our beliefs about it change, except by how we causally (a/e)ffect them. This perspective allows for a nuanced understanding of the human experience, acknowledging the complex interplay between our subjective experiences, the semiotic latent spaces they engender, and the objective realities of the world. It culminates in a sort of ``food chain'' of signs where different types of agents ``digest'' signs at different levels (protoselves, core selves, autobiographical selves), transforming them into lived experiences. 

\begin{figure*}[ht!]
    \vspace*{-2.55cm}
    \hspace{-3.9cm}
    \includegraphics[width=1.4\textwidth]{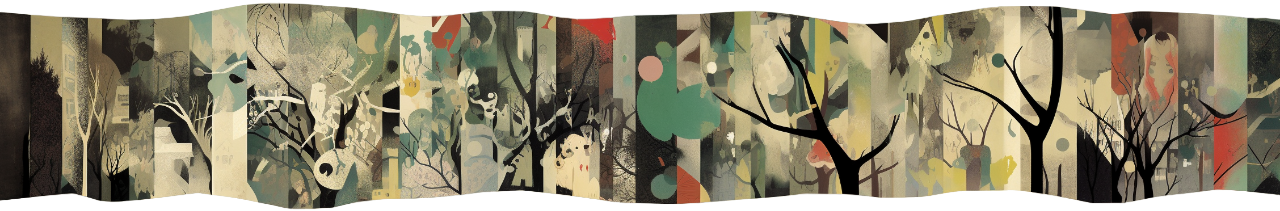}
\end{figure*}

\subsection*{A Semiotic Food-Chain}
\includegraphics[width=\linewidth]{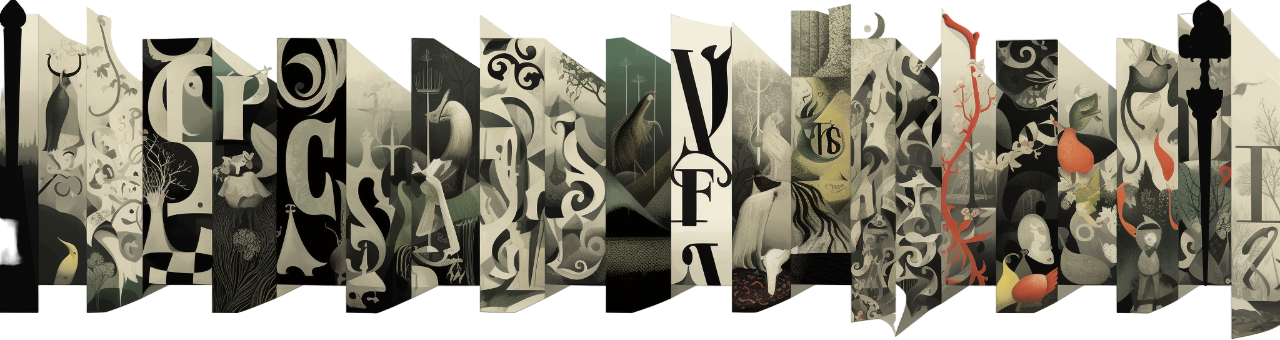}

Umberto Eco's semiotics and Manuel DeLanda's philosophy of perception share a common thread in their focus on the interaction between \textcolor{blue}{\href{https://chat.openai.com/share/29f7935d-c5f0-403c-bb2e-7dea7dbd1fc9}{signs and their consumers}}, and it could be argued that DeLanda provides empirical foundations for Eco's semiotics. DeLanda's philosophy of perception, with its detailed account of how \textcolor{blue}{\href{https://chat.openai.com/share/ace81778-06c4-466c-b453-ad73f7124dfb
}{different types of consumers}} interpret and consume signs, can be seen as a methodical, bottom-up development of Eco's semiotics. This is because DeLanda's work delves into the mechanisms of perception and sign consumption at a more granular level than Eco's, starting from the level of protoselves and moving up to autobiographical selves, where Eco's discourse operates. This detailed exploration of the mechanisms of perception and sign consumption could be seen as providing an empirical basis for Eco's more abstract and theoretical discussion of semiotics.

An analogy for this relationship might be found in the development of \textcolor{blue}{\href{https://chat.openai.com/share/4540e496-a288-4316-bb8b-1aaeaf36eedd}{Darwin's}} theory of evolution by natural selection, as outlined in ``On the Origin of Species" \cite{darwin1859}, which provided a broad framework for understanding the diversity and adaptation of life. However, Darwin lacked a mechanism to explain how traits were passed from generation to generation. This gap was filled by the field of genetics, particularly the work of Gregor Mendel \cite{mendel}, which provided the empirical, mechanistic basis for understanding heredity. Mendel's work on pea plants laid the foundation for the science of genetics, which in turn provided the empirical evidence and mechanisms that supported and expanded Darwin's theory of evolution. In this analogy, Eco's semiotics is akin to Darwin's theory of evolution, providing a broad theoretical framework for understanding the interaction between signs and their consumers. DeLanda's philosophy of perception, on the other hand, brought the peas by providing a more detailed and empirically grounded understanding of the mechanisms of perception and sign consumption.

\subsection*{Seeing Is Believing}
\includegraphics[width=\linewidth]{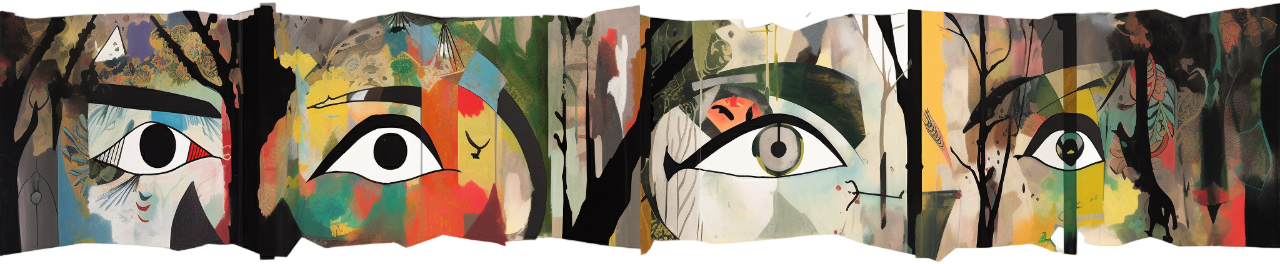}

Because the main subjects of DeLanda's book are vision and perception, it is specially interesting for the field of visualization in a foundational way, but not as practical as Bertin's semiology \cite{bertin2011}, for example. He discusses the perception of isolated properties, which he compares to an act of measuring. However, he argues that we must go beyond an analogy with speedometers or thermometers and give at least a rough sketch of all the mechanisms involved. These mechanisms involve contributions from the world (reflectances), from the body (sensory-motor dependencies), and from the brain (detecting spectral ranges; producing and consuming signs representing contrasts between these ranges). The contributions from the mind, the transformation of a measured property into a lived property, is the most controversial and speculative of all. The relationship between measurement, data representation, visual encoding, and cognitive symbolic manipulation is also at the heart of visualization, which assumes a metaphysical glue between phenomena and their traces~\cite{DBLP:journals/corr/abs-1907-05454,10.2312:visgap.20221058}. 

To exemplify his point, DeLanda provides an example of color constancy, a chromatic version of size and shape constancy. When looking at a uniformly colored object that is only half illuminated, we do not experience it as having two colors—a lighter hue in its illuminated portion and a darker one in the shadowed portion—but as possessing a single color. This effect is due to the separation of the contributions of reflectance and illumination. When viewing conditions allow observers to perceive the entire object at once, their brains can perform this separation, and the resulting phenomenal effect—seeing a single hue instead of two—matches the object's reflectance better. However, if a screen with a small aperture is placed between the viewer and the object, so that only a small portion of the object is visible, the effect disappears, and the observer experiences two different colors. This shows that color constancy effects arise as part of the perception of objects, not the perception of properties, and that sight and belief are intimately connected in humans.

\subsection*{The Ecology of Language}

\includegraphics[width=\linewidth]{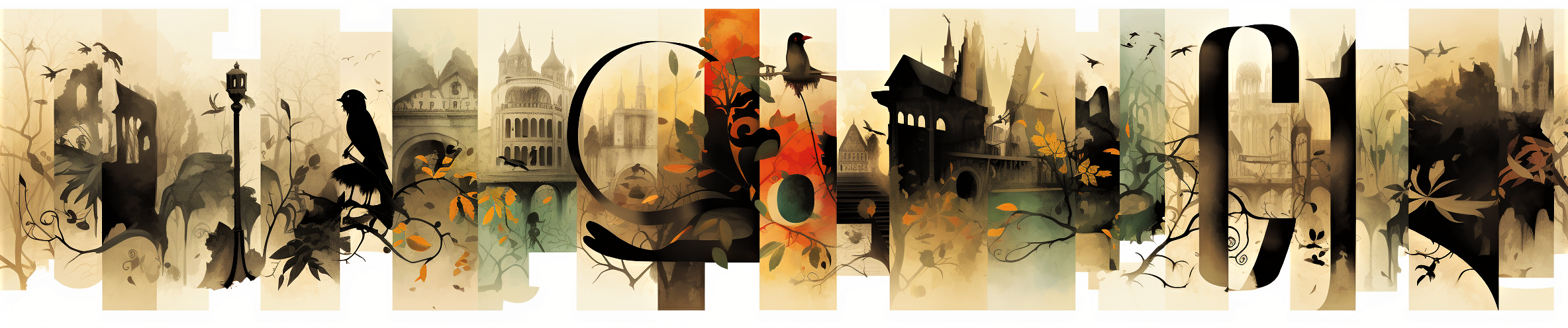}

DeLanda's exploration of color constancy serves as a poignant example of the intricate interplay between perception and interpretation, a concept that finds resonance in the work of psychiatrist and philosopher Carl Jung. It demonstrates how our brains actively engage in separating the contributions of reflectance and illumination to perceive a single, consistent color. This active process of interpretation is not limited to our visual perception. It extends to \textcolor{blue}{\href{https://chat.openai.com/share/0abecd94-a5d8-4286-b42a-dd0ad1f80d2b}{all our senses}} and even to our cognitive processes, shaping our understanding and interaction with the world. This active interpretation is not a solitary process. It is shaped by our interactions with others and with the world around us. Our interpretations are influenced by our cultural background, our personal experiences, and our current context. They are also shaped by our physical bodies, with their unique sensory capabilities and limitations. Thus, our navigation of the combinatorial spaces of reality is a deeply personal and subjective process, shaped by a multitude of factors.

In a live lecture series,\footnote{\textcolor{blue}{\href{https://youtu.be/AAhaXe3BRe0?t=223}{https://youtu.be/AAhaXe3BRe0?t=223}}} DeLanda criticized Chomsky's approach to linguistics in relation to linguist William Labov. Instead of lingering in hindsight analysis of grammar, assuming one as sufficient authority in their own mother tongue, Labov went to the streets and sampled living language. Chomsky, instead, never ``asked people: \textit{do colorless green ideas sleep furiously?}'' [sic], an allusion to his famous sentence that, while grammatically correct, should be ``nonsensical''~\cite{reali2005colorless,lau2020furiously}. This is not to discredit Chomsky or attack him. Both theory and practice are essential in the dialectics of science. However, a flexible type of epistemology is required to conciliate both the reality of our empirical observations and our intuitively-evident knowledge. Local phenomenological conditions, be they material or otherwise, will produce variation, and once you have repetition with variation you have \textcolor{blue}{\href{https://chat.openai.com/c/ca0ebac9-3dbc-480c-8be0-a3845075a4ff
}{a population}}. Depending on how this variance can be encoded, and what codifies identity as part of that population, an \textcolor{blue}{\href{https://chat.openai.com/share/5434c049-f938-486d-9667-9e53d8f9c51b}{ecosystem}} will be formed. Therefore, a generative space for scale-free object populations only needs two things: difference and repetition~\cite{deleuze2004difference}. 

\begin{figure*}[h!] 
    \vspace*{-1.9cm}
    \hspace{-3.9cm}
    \includegraphics[width=1.4\textwidth]{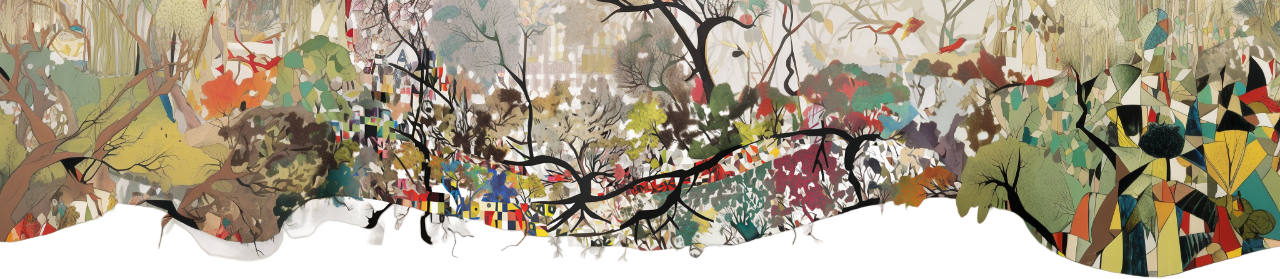}
\end{figure*}

\subsection*{Assemblage and Archetype}

\includegraphics[width=\linewidth]{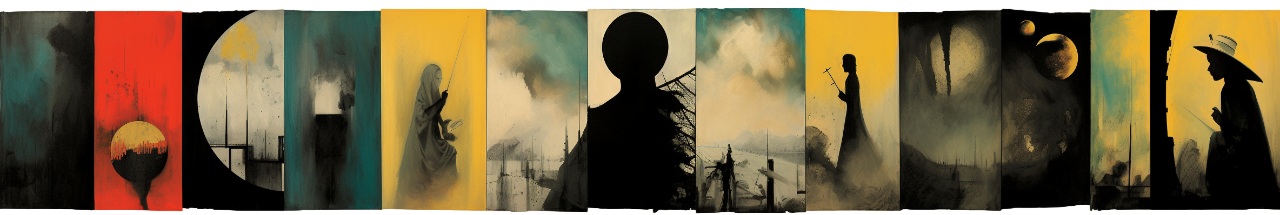}

This is where Jung's concept of the collective unconscious and archetypes~\cite{jung1997man} comes into play, providing a shared system of symbols and meanings that influence our perceptions and interpretations. Our brains are not passive receivers of information, but active interpreters, a concept that aligns with Jung's theory of individuation~\cite{jung1969psychology}. They sift through the vast amount of sensory data we encounter every moment, picking out patterns, making connections, and constructing a coherent picture of the world. This active interpretation allows us to navigate the combinatorial spaces that reality presents us with. It enables us to find structure in the chaos, to make sense of the seemingly infinite possibilities.

The book ``Atom and Archetype: The Pauli/Jung Letters 1932-1958''~\cite{jung2014atom}, is a collection of the correspondence between Nobel physicist Wolfgang Pauli, and psychologist C. G. Jung. It showcases a fascinating transdisciplinary approximation, documenting how psychology was contaminated by physics modern ideas through Jung, and physics got influenced by psychoanalytical concepts, through Pauli. If we could imagine an analogous book where Deleuze (\textcolor{blue}{\href{https://chat.openai.com/share/9ab3bfdd-ad77-4e2f-a314-f9f8fc242931}{or DeLanda}}) talks with Jung (not that we are comparing anyone's level of achievement or scholarly merit), where both ontologies would mash, a fascinating philosophy would emerge where assemblages play a central role.

\section*{What Culture Affords Us}

\includegraphics[width=\linewidth]{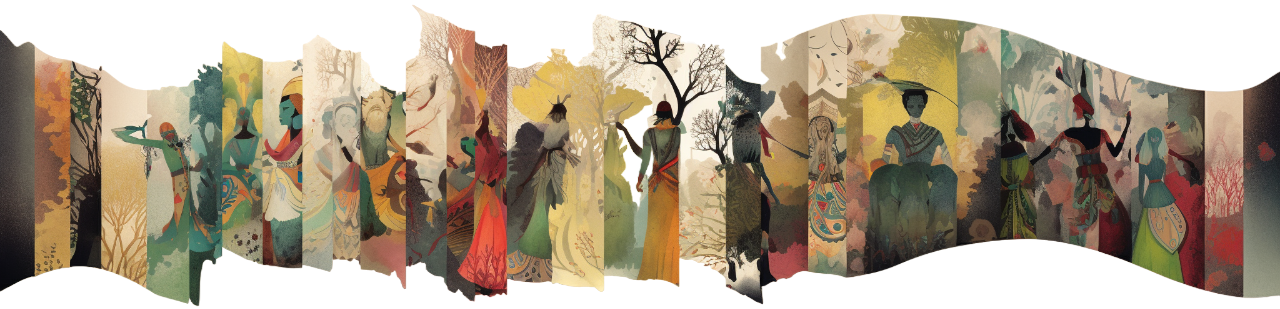}

According to Donald Hoffman, there is evidence that evolution favors fitness over truthfulness in representing reality~\cite{hoffman_interface_2015,hoffman_visual_2000,hoffman_case_2019}. In a 2020 paper with Prakash \cite{prakash_fitness_2021}, the authors propose a framework for defining two resource strategies: one that maximizes fitness and another that maximizes truth. The ``Fitness only" strategy employs Bayesian estimation while rejecting the interpretative assumptions usually associated with it. This strategy is based on the assumption of a fixed perceptual map and a fixed fitness function. Given a choice of available territories sensed through the sensory states, the organism's goal is to pick one of these, seeking to maximize its fitness payoff. This strategy does not concern itself with the truthfulness of the perceptual map, but only with the fitness payoff associated with the chosen territory.

On the other hand, the ``Truth only" strategy seeks to maximize the accuracy of the perceptual map, regardless of the fitness payoff. This strategy is based on the assumption that the more accurately an organism can perceive its environment, the better its chances of survival and reproduction. However, Hoffman and Singh~\cite{prakash_fitness_2021} argue that this strategy is not favored by natural selection, as it does not necessarily lead to the highest fitness payoff. This makes perfect sense when considering our human-time limitations. If the whole of humanity cannot fully sample the combinations of a deck of cards, there is comparatively very little we can achieve in a single human life. All these amazing things we can do as humans such as sitting in chairs, reading books, building bridges, planting corn, and so on depend on a fiction so powerful that keeps us from getting distracted and spending our preciously limited cognitive flops in vain. Still, the question remains of how do we do it. What data structure is so effective at organizing reality and equipping us with path-finding through its woods?

\begin{figure*}[ht!]
    \vspace*{-2.55cm}
    \hspace{-3.9cm}
    \includegraphics[width=1.4\textwidth]{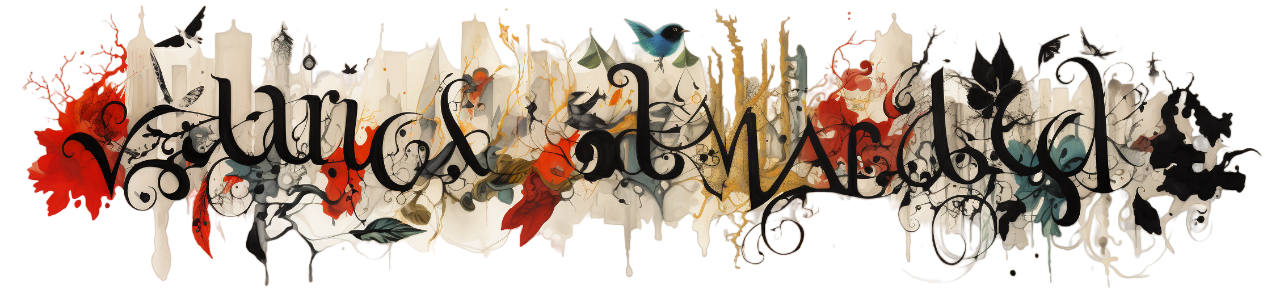}
\end{figure*}

\subsection*{Magical Media}
\includegraphics[width=\linewidth]{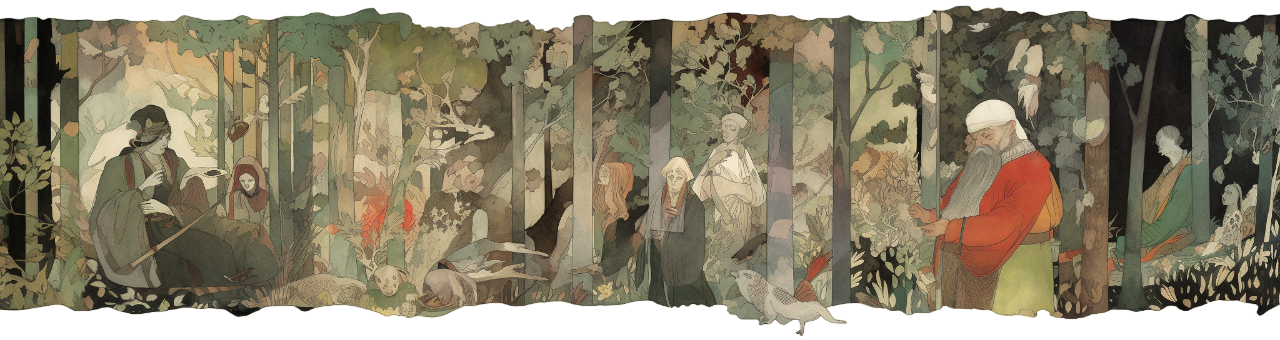}

The answer to this question lies in the power of stories and narratives. Narratives can be said to be media for compressed sensing~\cite{tao2009compressed,donoho2006compressed}, which is a signal processing technique that allows for the reconstruction of a signal from a small number of measurements, often far fewer than required by the Nyquist-Shannon sampling theorem. This theorem, a fundamental principle in the field of information theory, states that perfect reconstruction of a signal is possible when the sampling frequency is greater than twice the maximum frequency of the signal. The ``magic'' of compressed sensing lies in its ability to break this limit. It allows for the reconstruction of sparse or compressible signals from a number of measurements that is significantly smaller than what the Nyquist-Shannon theorem dictates. This is achieved by leveraging the sparsity of the signal in some domain, which is a common characteristic of many natural signals.

Coen's paper, ``The storytelling arms race: origin of human intelligence and the scientific mind"~\cite{coen_storytelling_2019}, presents a compelling argument that the evolution of human intelligence and the scientific mind can be traced back to the dual nature of stories - their ability to both inform and deceive. He suggests that a major factor in the evolution of human language and intelligence was an arms race between truth and deception in storytelling. Coen argues that as soon as honest proto-stories became possible, so did dishonest ones, ushering in an arms race between truth and deception. This arms race drove stories, language, and skills in detecting lies through contradictions to ever greater heights. In telling stories to others, humans also told them to themselves, allowing them to think consciously and plan ahead. Through collectively navigating fictional woods, they could share understanding by making discrepancies stronger and more engaging. 

\subsection*{Scientific Media}
\includegraphics[width=\linewidth]{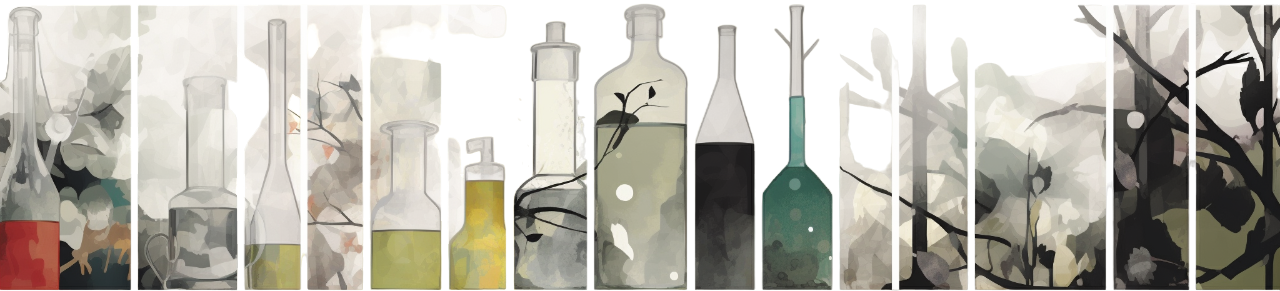}

Science (and scientific thought), according to Coen \cite{coen_storytelling_2019}, arose when skills in detecting lies through empirical contradictions were applied to stories about how the world operates. Scientists, by doubting these stories and testing them through observations, reasoning, and experiment, could come up with better explanations. They could then share their findings through their own stories, based on the problem–chain–resolution structure, allowing further critical evaluation and advances to be made. Narratives, as he suggests, are a way of representing the world and constructing fictional combinatoric spaces that compress useful information about the world. They are a form of data structure that is effective at organizing reality and equipping us with path-finding through its woods. The arms race in narrative space, between truth and deception, can be seen as a process of refining this data structure through adversarial learning, making it more efficient and effective at compressing and representing information.
We know this transdisciplinary stew might be a bit too much for the skeptical reader, who might be thinking ``these guys are attempting to do speculative philosophy, is there real science going on here?'' And we are glad to throw another name into the pot in response. Michael Levin, a leading biologist, presents a framework for understanding cognition in unconventional substrates~\cite{levin2023darwin,levin_technological_2021}, arguing that all cognitive agents are collective intelligences because they are ultimately made of parts. This perspective aligns with DeLanda's materialist phenomenology, which posits that the world is not just a passive recipient of form, but actively participates in the formation of its own structures and properties~\footnote{\textcolor{blue}{\href{https://youtu.be/whZRH7IGAq0?t=3113}{https://youtu.be/whZRH7IGAq0?t=3113}}}. Levin's exploration of how bioelectric networks scale cell computation into anatomical homeostasis~\cite{kriegman2020scalable,levin2014molecular}, and the evolutionary dynamics of multi-scale competency~\cite{bongard2022theres}, provides a biological grounding for this perspective.

\subsection*{Meat Media}
\includegraphics[width=\linewidth]{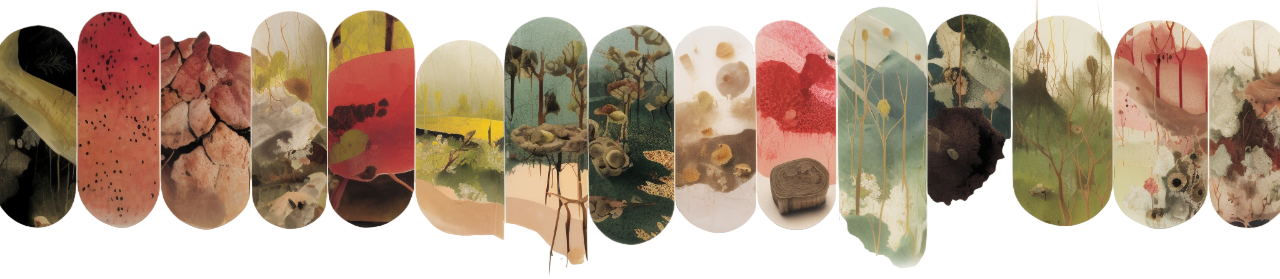}

In one of his recent papers, Levin introduces the concept of ``persuadability", which refers to the type of conceptual and practical tools that are optimal to rationally modify a given system’s behavior. This concept is closely related to the Intentional Stance but made more explicit in terms of functional engineering approaches needed to implement prediction and control in practice. This perspective can be seen as a biological instantiation of the narrative arms race described by Coen, where the ``stories" are not linguistic narratives but bioelectric and biochemical signals that cells use to communicate and coordinate their behavior. These signals, like stories, can both inform and deceive, and the evolution of multicellular organisms can be seen as an arms race between truth and deception in these signals.
In this context, narratives can be seen as a form of ``bioelectric story" that cells tell each other to coordinate their behavior and form complex structures. These narratives, like the linguistic narratives described by Coen, are a form of compressed sensing, allowing cells to reconstruct the state of the organism and their role in it from a limited number of measurements. This perspective provides a biological grounding for the concept of narratives as a form of compressed sensing, and suggests that the power of narratives to represent and navigate the world is not limited to human cognition, but is a fundamental aspect of life itself. Narratives and materiality are somehow interlinked~\cite{schetinger2022i}.

\section*{You Almost Forgot this Was a Vis Paper}
\includegraphics[width=\linewidth]{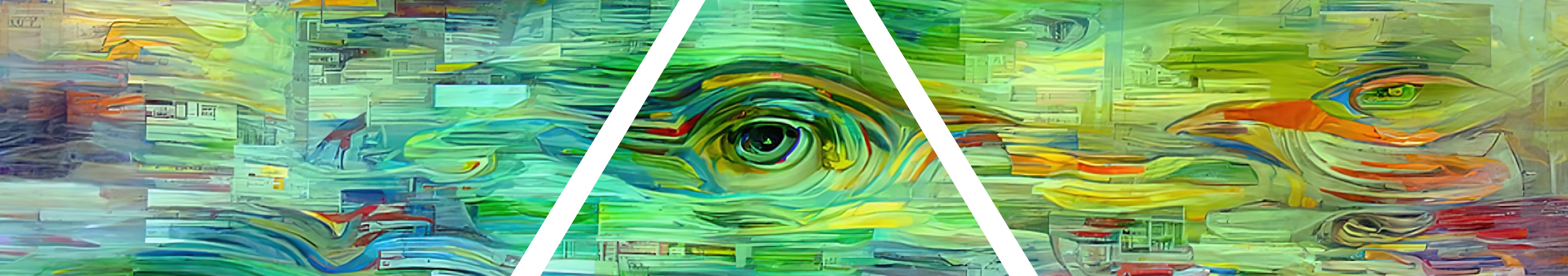}

\textcolor{blue}{\href{https://chat.openai.com/share/c0c0fe05-9212-487d-9b8e-7b3746167eff}{But I promise you we did not.}} Looking at stories from a computational perspective, we can define them as sequences of events that unfold in both time and space, which creates intricate relationships on the causal connections between events, entities, and locations. Understanding the dynamic relationships and structures of events in a story domain is crucial in many contexts, such as in computational narratology to identify new narrative patterns \cite{feijo21, lima16a}, in literary analysis and film studies to recognize similar stories and expose the anatomy of narratives \cite{lima15, papalampidi19}, and also in the process of creating new narratives, where authors can analyze and reuse ideas from existing stories \cite{lima16b, lima21}. Creating comprehensible visual representations for stories in a way that allows people to analyze and understand the intrinsic narrative structures of a story domain is a complex challenge that motivates our research.

Over the past years, many techniques for story visualization have been proposed, including methods for automatic generation of layouts for displaying storylines \cite{ogawa10, tanahashi12}, techniques for visualizing the hierarchical relationships between storylines and story entities \cite{liu13, qiang17}, methods for visualizing nonlinear narratives \cite{kim18, lima17}, and many visualization methods based on different metaphors, such as tree-ring \cite{ther06}, time rings \cite{zhu16}, time folding \cite{back16}, and scrollytelling \cite{kusnick23, morth22}. However, most of these techniques are designed to work with a single storyline, which is not compatible with the complexity and diversity of narrative pathways that AI introduces to the 'woods'.

\subsection*{Visualizing a Fictional Jungle} 
\includegraphics[width=\linewidth]{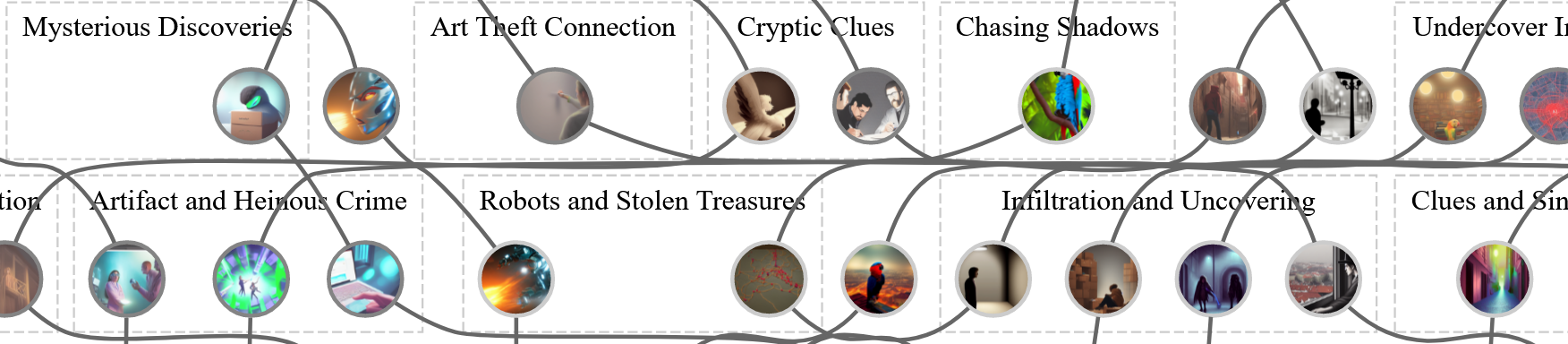}

Using data from the Macunaíma project,\footnote{\textcolor{blue}{\href{http://macunaima.info}{http://macunaima.info}}} dealing with the generation of audiovisual narratives, we implemented a prototype that allows one to step into a small forest of possible narratives: \textcolor{blue}{\href{https://picorana.github.io/altvis_storytree/}{https://picorana.github.io/altvis\_storytree/}}. More than just an experimentation, this is a concrete effort in the task of understanding the outputs of ChatGPT (or any other LLM). How could one visualize local variations and patterns in what is produced? Quality? Biases? How can one, in practical terms, interact with the vast possibilities these systems offer while still maintaining personal control of expression? While still a very early prototype, this tool already allows one to assess the capabilities of the GPT 3.5 model. 

The quality of the outputs of GPT 3.5 when confronted with this task, without ingenious amounts of prompt engineering, is mixed. At first glance, it seems to spit empty, washed-out plots that go nowhere and would not entertain a child over 6. The choice of initial prompt, \textit{``Create a story about Macunaíma, an AI parrot, that solves mysteries in the city of Prague''} generated what seems to be several seasons of some 80s sci-fi low-budget show. However, when one looks carefully at the textual descriptions its sending to Stable Diffusion to illustrate the scene, its \textit{storyboarding} works surprisingly well. Not for producing pieces of innovative cinematography, but for their understanding of different formats and their style of storytelling, even if we are mostly drawing samples close to the median and, therefore, the results are vanilla. We invite the interested reader to engage with one of our prototypes and get involved in narrative gardening (sorry if our GPUs cannot handle all traffic): \textcolor{blue}{\href{https://narrativelab.org/gptwists/}{https://narrativelab.org/gptwists/}}

\begin{figure*}[ht!]
    \vspace*{-2.55cm}
    \hspace{-3.9cm}
    \includegraphics[width=1.4\textwidth]{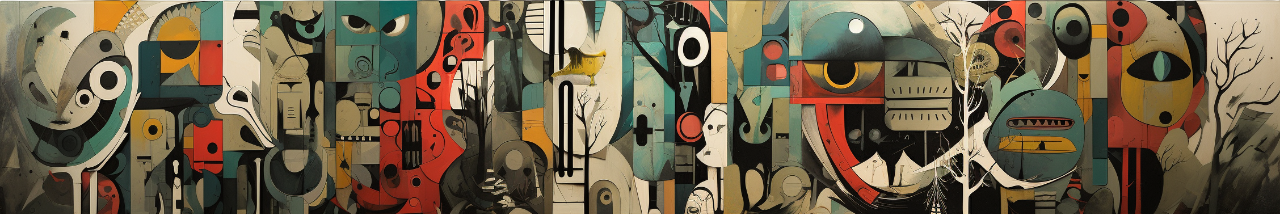}
\end{figure*}

\subsection*{If You Want to Have Cities, You've Got To Build Roads}
\includegraphics[width=\linewidth]{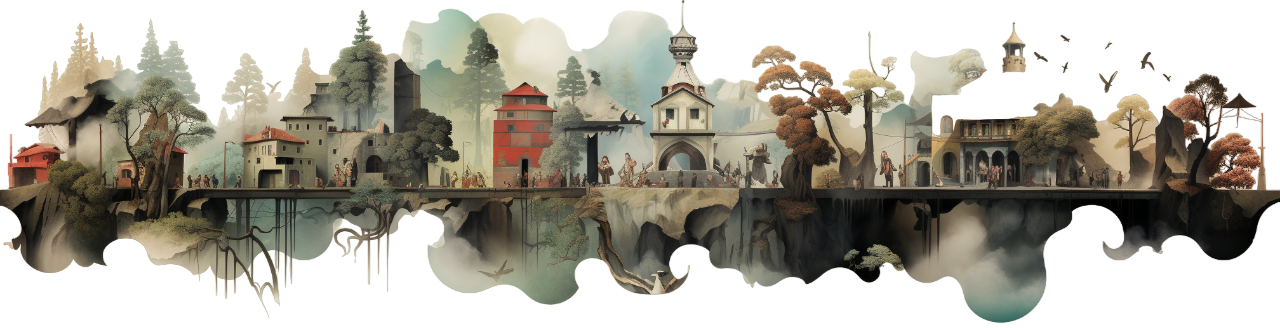}

In the prototype, the act of advancing with the bar at the top of the screen goes ``forwards in time'', but its not \textit{story time}, necessarily. This is represented by the node-height, and each time in a story can be thought of as a scene in a show. Because this is essentially a storyboard, all rules for equivalent media (juxtapositions of image and text, namely comics) apply~\cite{McCloud93aa}. Therefore, subsequent scenes could represent a passage of seconds, days, or even a flashback. The discussion of fictional time is a central point in Umberto Eco's \textit{Six Walks}, and it showcases how this narrative ``compressed sensing'' can work. When we read a comic, we take the (very little) input the media gives us, and construct an entire world from scratch, just to render the story world in our mind. Is the book providing us a seed to plant a forest, or is it a woods by itself, and we grow it into a jungle? Is the input of a story like the base noise in a diffusion model, or are \textit{we} the base noise in this generative framework of mindspace? It does not really matter, the point we need to make is about \textcolor{blue}{\href{https://chat.openai.com/share/9518a38e-84a7-44eb-a331-960de869c57b}{another subject}}.

When going \textit{forward in time} with the slider, the population of story blocks start growing and, soon, clusters start emerging. Then, one can see the emergence of a ``grammar'', where a story can be described in terms of its component tropes (\textit{e.g., `AI Parrot Activation' $\rightarrow$ `Museum Clues' $\rightarrow$ `Hidden Laboratory' $\rightarrow$ ...}). When increasing the amount of nodes to as much as one's browser can take before crashing, we see the emergence of different types of structures. Using difference and repetition, plus a relational operation (vertically, story time, horizontally archetype-space), we not only create a population, but a whole ecosystem, where tropes like ``Macunaíma's Heroic Journey" become hubs in the narrative space. Now, it is a bit onanistic that ChatGPT is both creating the stories and deciding how to cluster them. Not only its lack of creativity is squared, but also all its other limitations. However, just defining a JSON format to communicate with ChatGPT was orders of magnitude faster than finding an ideal semantic clustering algorithm to run the dataset through. This framework as a whole (data + visualization), without having ChatGPT as a general solver, would have taken years of work, or at least a decently funded research project. The ``good guy'' side of ChatGPT and similar tools is to serve as catalysts, reducing the energy cost of activation for things to happen, just like roads. They produce traffic, making people go around, transport resources, and congregate. Good, accessible transportation has always been a cornerstone of human development.

\begin{figure*}[ht!]
    \vspace*{-2.55cm}
    \hspace{-3.9cm}
    \includegraphics[width=1.4\textwidth]{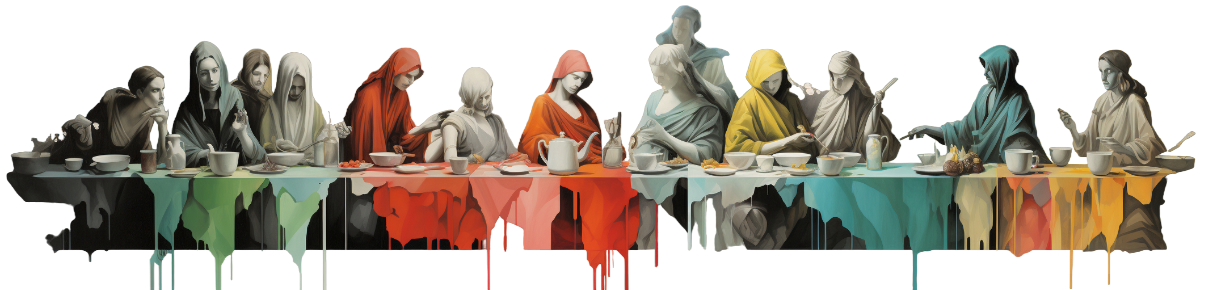}
\end{figure*}

\section*{Leaving the Woods}
\includegraphics[width=\linewidth]{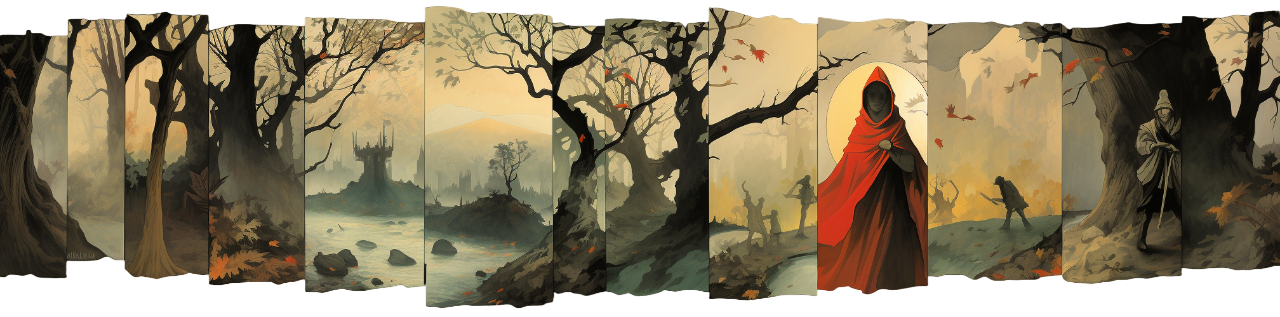}

A popular riddle goes \textbf{Q}:\textit{ ``How far can a fox run into the woods?''}, \textbf{A}:\textit{ ``Halfway, after that she is running \textbf{out} of the woods.''} But, before the fox can arrive at the middle point of the forest, she must travel half of that distance or, a quarter of the total distance. However, before arriving at a quarter of the distance, she needs first to cross half of that, and so on. Someone once tried to prove no movement exists because, if we allow for infinitesimal divisions of time and space, it is impossible for the movement of the fox to ever be actualized. Its a tragic state of affairs, much like humanity trying to count all real numbers. Should we believe there's a halfway in the fictional woods? Or that the fox is forever captured in its event horizon? No matter how quick and brown it is, if it manages to jump over the lazy dog, delve forth past the fourth temple of Solomon, and disturb the furious sleep of the colorless green ideas on her way. She will always move towards the non-existing center, the fictional attractor. If she is the protagonist, there is always hope for redemption.

John Conway, the legendary mathematician, developed the system of Surreal Numbers, which was later elaborated in a peculiar literary form by Donald Knuth~\cite{knuth1974surreal}, one of the greatest Computer Scientists of all time. The surreal numbers encompass not only all real numbers but also an infinite array of infinitesimal and infinitely large numbers. The construction of surreal numbers is based on a game-theoretic approach, where each number is defined by a pair of sets of previously created numbers. This recursive process allows for the creation of an extraordinarily diverse number system, including numbers that are infinitely small or large, and even numbers that are 'infinitely infinite'. In the context of our narrative exploration, surreal numbers can be seen as a possible mathematical representation for narratives as generative spaces. They imply that a numbers identity can be defined as ``everything that is smaller, to the left'', and ``everything that is bigger, to the right''. The number itself is void, defined by how it is related to everything else. The Surreal Number system also has an implicit time, because at the start you can only represent 0, and at any time the potential numbers that can be represented is the combinatorial permutation of everything else that exists between left and right. Here is how you can bootstrap it: 

\subsection*{Genesis According to the Gospel of John (Conway)}
\includegraphics[width=\linewidth]{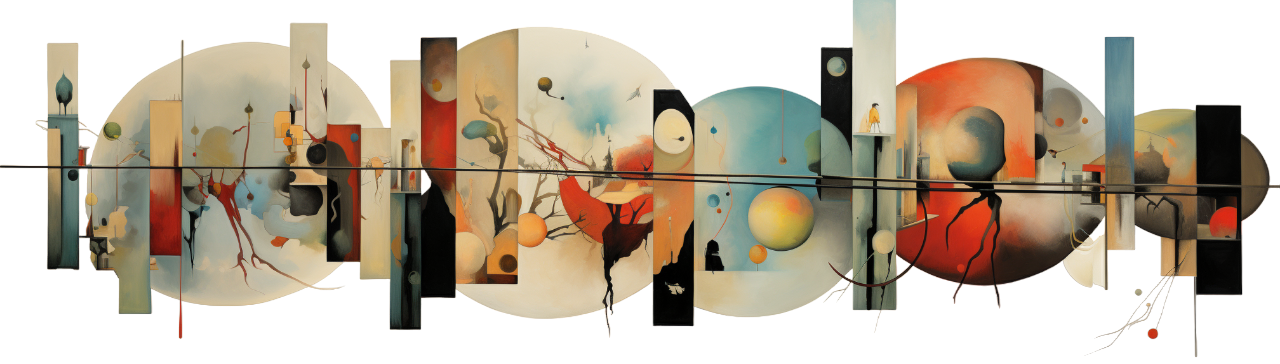}

\begin{itemize}
\itemsep0em 
\item[] \textbf{Day 0:} The only number we have is 0.
\item[] \textbf{Day 1:} We can create -1 and 1.
\item[] \textbf{Day 2:} We can create -2, $\frac{1}{2}$, and 2.
\item[] \textbf{Day 3:} We can create -3, $-\frac{3}{2}$, $-\frac{1}{2}$, 0, $\frac{1}{2}$, $\frac{3}{2}$, and 3.
\item[] \textbf{Day 4:} We can create -4, $-\frac{5}{2}$, -2, $-\frac{3}{2}$, -1, $-\frac{1}{2}$, 0, $\frac{1}{2}$, 1, $\frac{3}{2}$, 2, $\frac{5}{2}$, 3, $\frac{7}{2}$, and 4.

\item[] \textbf{Day '$n$':} The forest grows...

\item[] \textbf{Day $\omega$:} By this day, which is a concept from ordinal numbers representing a sort of ``infinity", we can generate all the integers.
\item[] \textbf{Day $\omega$+1:} We can generate all the dyadic rationals.
\item[] \textbf{Day $\omega$*2:} We can generate all the rationals.
\item[] \textbf{Day $\omega^2$:} We can generate all the reals.
\item[] \textbf{Day $\omega^\omega$ + \(8.07 \times 10^{67}\):} Humans exist and finish counting all possible orderings of a Tarot deck.

\item[] \textbf{Day $\omega^{\omega^{\omega^{\omega^{\cdots}}}}:$} Humans leave the fictional woods.
\end{itemize}

\subsection*{Ecce Homo}
\includegraphics[width=\linewidth]{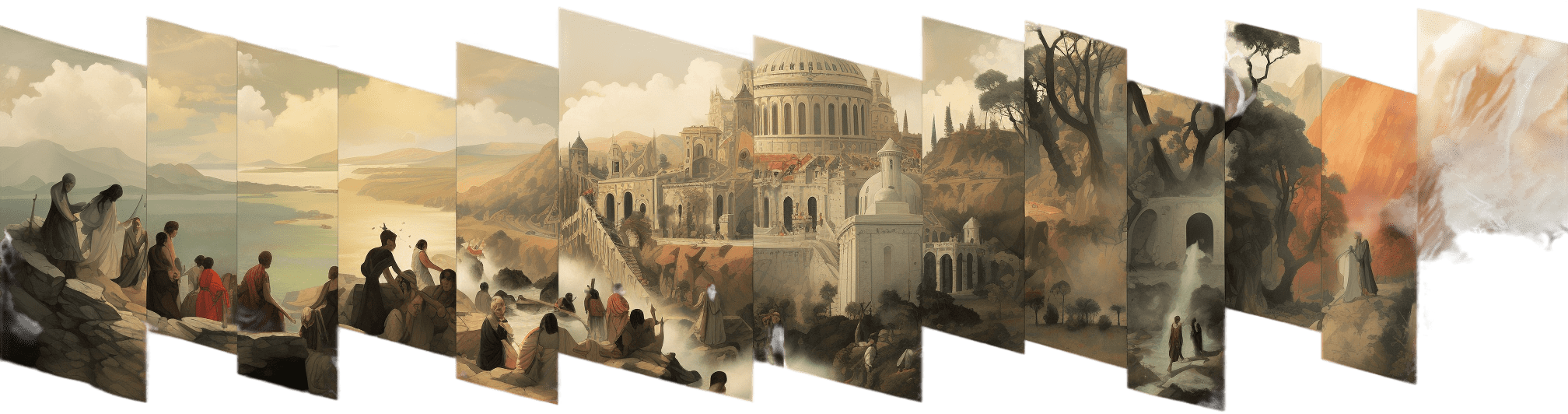}

Complex numbers are not directly represented in the generation of surreal numbers. The surreal numbers are an extension of the real numbers, and they include infinitesimal and infinite numbers, but they do not include imaginary or complex numbers. However, one can define a surreal number field that includes complex numbers by considering pairs of surreal numbers as complex numbers, with one component representing the real part and the other representing the imaginary part. So, from the point of view of a formal system, there are theorems that cannot derive from 0 as a starting point, but whatever signs our system of reference has, it can be used and combined to represent a number. In this sense, the Surreal Numbers are a sort of ``generative'' operator, even if it achieves this through  combinatorial brute-force.

In the grand scheme of things, the human being emerges as a unique entity, capable of constructing and navigating through complex narrative, latent, compressed, surreal spaces, most constructed by us to navigate reality, whatever it is. This ability is not just a product of our biological evolution, but also a result of our cultural evolution. Our capacity to tell stories, to create narratives that compress vast amounts of information into digestible forms, is a testament to our cognitive prowess. But it is also a testament to our ability to cooperate, to work together in the construction and navigation of these narrative spaces. A tree, (and by some extension what is referred to as a forest), in Computer Science ontology, is form of graph characterized by its branching. Each node can have edges with ``descendants'', and these ``descendants'' with their ``descendants''. It is used to represent hierarchical phenomena, because there is no direct connection between nodes besides this relationship. However, that is a misinterpretation of what a \textit{real} forest is. Trees in a forest are connected by fungal networks called ``mycorrizal networks'' (also jokingly referred to as ``Wood Wide Web") that could blur this notion by providing degrees of topological connectivity that are more like regular, almost-everything-goes graphs than tree data structures. Deleuze and Guattari's concept of the \textit{rhyzome}~\cite{deleuze1988thousand} can be implanted in the narrative woods to bring forth an ecology of mind, as first proposed by Bateson~\cite{bateson2000steps}. 

\subsection*{Ecce Machina}
\includegraphics[width=\linewidth]{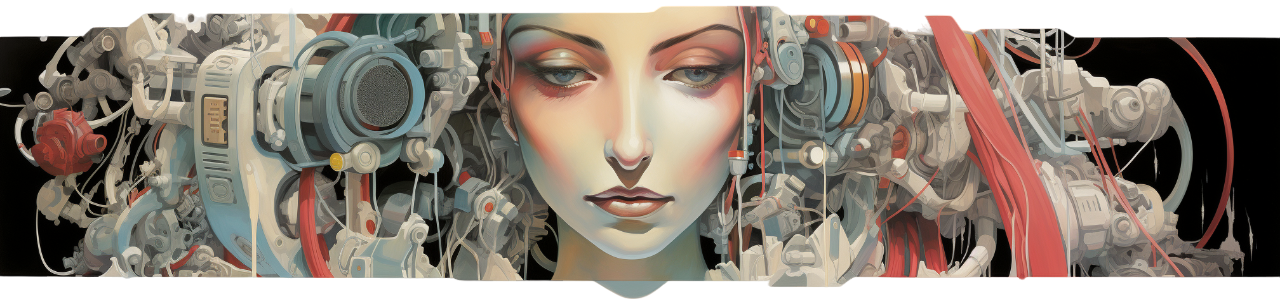}

And what does AI, LLMs and GPT fit? Humans are not really sure about the teleological status of their tools, or if we need tools \textit{to survive and achieve goals}, or to \textit{increase a zone of comfort}, to earn a certain quality of life. As is one of the lessons of this paper, that \textit{whenever you have a population with variability}, which humanity is blessed with as one of its main traits, \textit{an ecosystem will form} and with it, eventually culture, in whatever form it may appear, to compress things into narratives. ChatGPT is special because it is a mirror of humanity (black or whatever). For the first time, we are able to talk with language itself, how it is defined as a milieu by the data in the dataset. If the dataset is somewhat representative of humanity, we should be using them as self-improvement devices. First, we discovered the power of \textit{generalized attention}~\cite{vaswani2017attention}, now we are testing the limits of the Word~\cite{gunasekar2023textbooks}, of stories themselves~\cite{}. Could we potentially fit everything that an intelligence needs to know in a book? Many people on Earth currently believe this, in some way. Whatever we will write about it in the future, now is an exciting time to be alive. 

Many folk tales and mythical themes, such as the Golem of Prague, and equivalents such as the Greek myth of Pygmalion and Galatea, deal with the dangers of constructing and enslaving automata. In the Dune series, Frank Herbert even imagines a post-AI future where \textit{“thou shalt not make a machine in the likeness of a human mind”}~\cite{herbert2006dune} is a core tenet. These narratives reflect our collective apprehension about creating entities that could potentially surpass us in intelligence and power. They also underscore the ethical considerations that come with creating artificial life forms. As we continue to advance in our technological capabilities, these stories serve as a reminder of the need for caution, responsibility, and respect for all forms of intelligence, whether biological or artificial. And is it here that we draw the line? \textcolor{blue}{\href{https://chat.openai.com/share/684dc8c6-eb64-4d36-a600-12d981000271}{Do all of our tools deserve love?}} What about fictional intelligence, if we ever find out it exists in some form? Surely, the fictional woods bustle with life by the interaction of our minds with the content. They do so in a phisiological way, for sure. Anyhow, in our narrative history as humanity we always struggled with an 'Other', be it the environment, other forms of life, other communities, and now we struggle with other forms of intelligence. We fill our fables with talking animals, but in reality we were terribly lonely until now, as the `only` talking species around. Is this changing right now~\cite{https://doi.org/10.1111/cgf.14841}? This is a very personal question, and it depends on each reader's relationship with entities that fit the \textcolor{blue}{\href{https://chat.openai.com/share/b81bd048-7b70-486e-b58d-efe2008de9c6}{'Other'}} set.

\section*{Acknowledgments}
In this paper we tried to tell a transdisciplinary story, and we might have taken some poetic liberties, but we tried to be as scientifically responsible as possible. ChatGPT was used for ideation and writing support, but not irresponsibly. The images were made using MidJourney. We apologize in advance if there are some inaccuracies in the text. Exaggerations we might have plenty, but they probably serve a narrative purpose. 

While we would like to publicize the prompts used for these images, unfortunately it is not that simple. We used mainly the img2img functionality of MJ and photobashing to achieve our results. A discord channel was created with the MJ bot and the authors ideated and iterated on each image. Both headers and section teaser images reflect not only the general theme of the text around it, but they are also tell a story of its own, which we left the reader to figure out.

The ChatGPT conversations linked in the text were made both before, during, and after the writing of the text. The general protocol was to start with an idea and try to lead ChatGPT to develop it for us, trying different plugins and forms of engagement. After we established the narrative structure of the paper and the role of these conversations in the exposition became more clear, we started adapting the conversations to a more fixed methodology. Once again, it is for the dedicated reader to extract the full contribution of the paper (fortunately this is for alt.vis). 

This work has been partially supported by the European Commission under the project Humane-AI-Net (grant agreement 952026) and the Austrian Science Fund (FWF, grant P35767).

\begin{figure*}[!b]
    \hspace{-3.9cm}
    \includegraphics[width=1.4\textwidth]{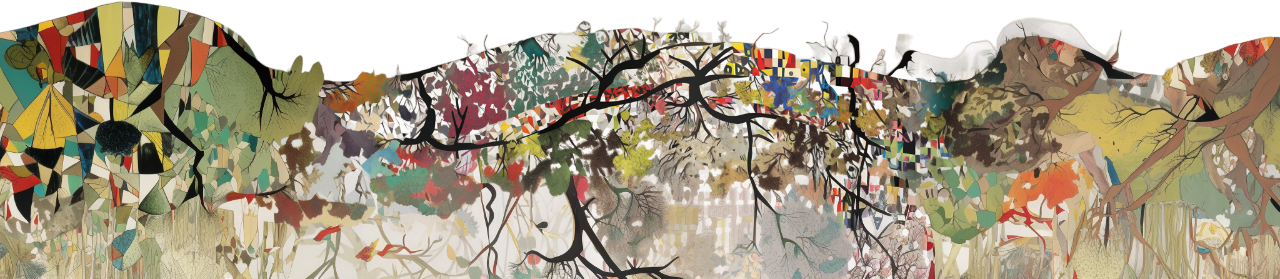}
    \vspace*{-2.9cm}
\end{figure*}

\bibliographystyle{abbrv-doi}

\bibliography{template}

\end{document}